\documentclass[aps,prl,twocolumn,groupedaddress,floatfix]{revtex4}  
\usepackage{graphicx}  
\usepackage{dcolumn}   
\usepackage{bm}        
\usepackage{amssymb}   
\usepackage[export]{adjustbox}

\usepackage{sidecap}
\usepackage{setspace}

\usepackage{color}

\usepackage{amsmath}

\begin{document}

\title{CEP-Controlled Molecular Dissociation by Ultrashort Chirped Laser Pulses}

\author{S. A. Karakas$^1$}
\author{P. Rosenberger$^{2,3}$}
\author{M. F. Ciappina$^{2,4}$}
\author{M. F. Kling$^{2,3}$}
\author{I. Yavuz$^1$}
\affiliation{\\}
\affiliation{$^1$Physics Department, Marmara University, 34722 Ziverbey, Istanbul, Turkey}
\affiliation{$^2$Max-Planck-Institut f\"ur Quantenoptik, Hans-Kopfermann-Stra{\ss}e 1, D-85748 Garching, Germany}
\affiliation{$^3$Department of Physics, Ludwig-Maximilians-Universit\"{a}t Munich, Am Coulombwall 1, D-85748 Garching, Germany}
\affiliation{$^4$Institute of Physics of the ASCR, ELI-Beamlines, Na Slovance 2, 182 21 Prague, Czech Republic}

\begin{abstract}
We demonstrate and characterize that a carrier-envelope-phase (CEP)-controlled ultrashort chirped field is an efficient and robust mechanism to modify the dissociation dynamics of molecular hydrogen. Different dissociation pathways are collectively induced and their interference  contribute to the kinetic energy release spectra. Chirping is able to efficiently manipulate the interferences of different dissociation pathways. We demonstrate a linear relationship between chirp and CEP-dependence, dissociation as well as directional electron localization. 
\end{abstract}

\maketitle
The advent and steady evolution of ultrashort intense laser sources and momentum imaging techniques, have placed the research in the fields of ultrafast strong-field-matter interaction in general and photodissociation and ionization processes in particular attention \cite{intro1, intro2, intro3, intro4, intro5, intro6, Kling246, intro8, intro9, intro10, intro11, intro12, intro13, intro14}. The simplest molecules available, i.e.,  the molecular hydrogen and its ionic companion, have been used as standard models to examine the underlying complex dynamics arising when they are illuminated by strong and short light fields. Particularly interesting phenomena, such as bond softening and above-threshold dissociation have been experimentally demonstrated successfully in these elementary molecular systems \cite{intro4, intro10, intro11, intro14, intro15, intro16}. Amongst the knobs available to manipulate and control these phenomena, laser peak intensity, carrier envelope phase, and pulse duration have predominantly been used \cite{Kling246, Alnaser_2017, Li_2017, Ibrahim_2018}. It is possible, however, to manipulate the temporal and spectral shape of the driven laser sources \cite{intro17}, thanks to the spectrally broad bandwidth of femtosecond pulses. The linear chirp, the linearly instantaneous central frequency change across the bandwidth of the pulse, configures the most simple example of such control. Pulses in which the lower frequencies come behind (ahead) are referred to as negatively (positively) chirped. More involved manipulation schemes can be used though, including nonlinear chirped pulses \cite{intro18} and multicolor laser sources \cite{intro19}. Applications of chirped pulse driving include; controlling atomic collisions~\cite{wright2007},  resonant high-order harmonic generation \cite{abdelrahman2018chirp} and above threshold ionization \cite{xu2010}, single attosecond pulse generation \cite{niu2009}, controlling the population transfer in ro-vibrational states~\cite{sarkar2008, plenge2009} and imaging coupled electron-nuclear dynamics~\cite{jelovina2018pump}. 
The theoretical and experimental complexity of them, however, make linear chirped pulses still the most suitable drivers for gaining systematic knowledge. In particular, the manipulation of bond hardening and selective bond dissociation and dissociative ionization using linearly chirped pulses were demonstrated \cite{intro8, jelovina2018pump, natan2012}. We should note, however, that none of these studies have specifically reported relevant effects in ultrashort pulses due to the chirp sign on the photodissociation of $H_2^+$. 

Chirping the laser pulse not only introduces polychromy with certain gradients, but also directly reduces the intensity by $I_0/(1 + \eta^2)$ and increases the temporal pulse bandwith $(1 + \eta^2)\tau_0$ due to group delay dispersion (GDD), where $I_0$ and $\tau_0$ are the intensity and duration of the transform-limited (TL) pulse, respectively, and $\eta$ is a unitless parameter to define the linear chirp. Reduction in intensity or ending up with a longer pulse is certainly undesirable for ultrashort laser pulse generation and for efficient molecular dissociation due to the fact that pulse-elongation via chirp makes the dissociation rate suppressed~\cite{Vibok2} as well as disrupts CEP-control, which are both crucial for directional control of molecular reactions. To overcome these issues in chirping, one can consider a phenomenological temporal-bandwidth-maintaining chirping scheme where the intensity and the duration are kept fixed for varying chirps. We anticipate that this concept allows more robust investigation of coherent control over the dissociation dynamics of molecules. 

\begin{figure}[t!]
  \includegraphics[width=\linewidth]{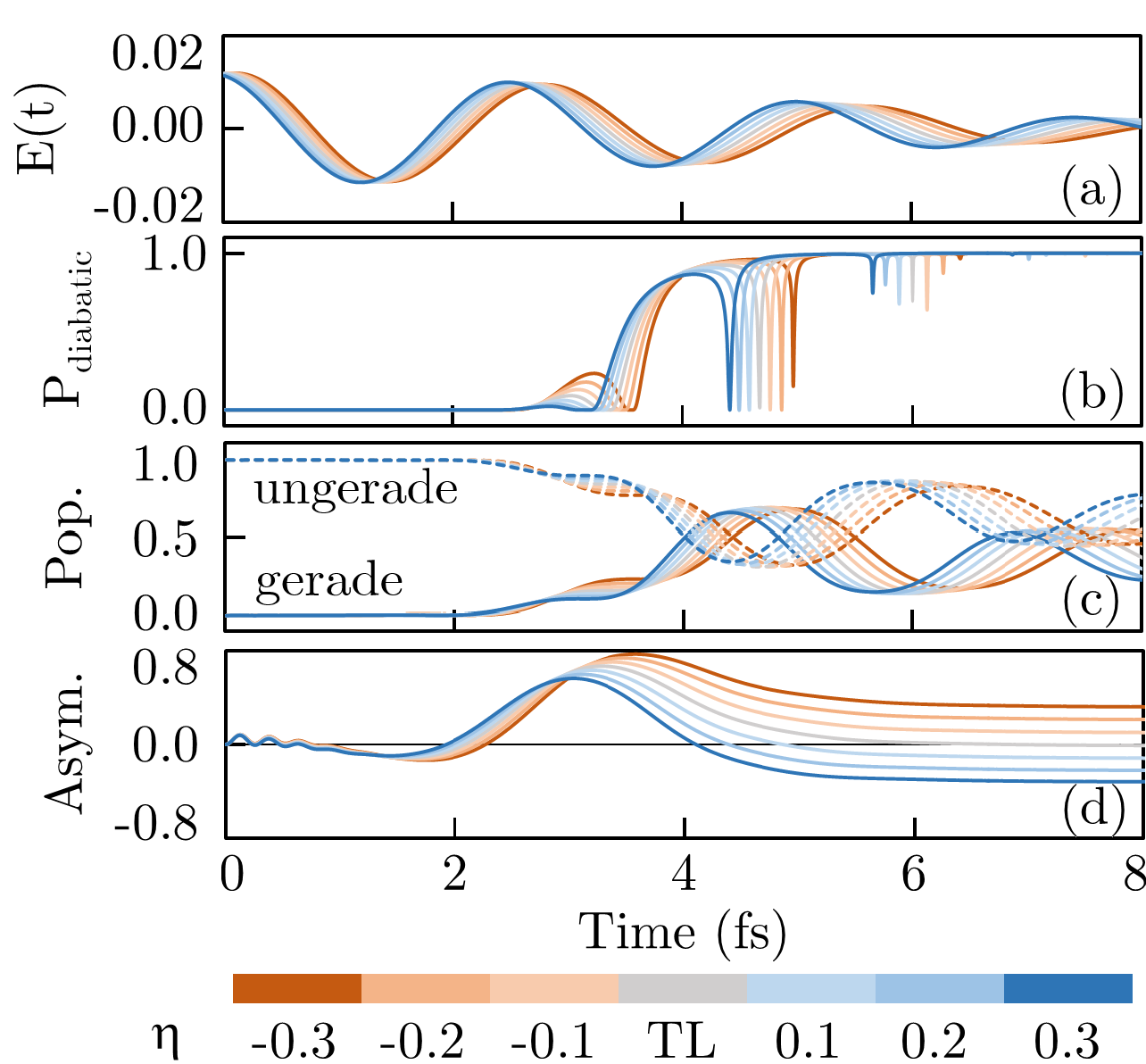}
  \caption{(color online) Chirp effect on the dissociation of $H_2^+$ in a two-level system. All grey(blue)[brown] lines for transform limited(positively perturbed)[negatively perturbed] pulses. (a) Temporal variation of the chirped pulse. $\phi_0=0$. (b) Landau-Zener probability from adiabatic to diabatic process. (c) Left(solid) and right(dashed) populations over the ungerade and gerade states. (d) Asymmetry change by chirp. It changes sign with the change in the sign of chirp. \\}
  \label{fig:two_level}
  \vspace{-8.0mm}
\end{figure}

We consider the chemical reaction $H_2 + \gamma \rightarrow H_2^+ +  \gamma\rightarrow H + p$, where $\gamma$ is the chirped field. Here, we assume that the cation is on-the-fly populated from the ionization of $H_2$ by the laser field. MO-ADK~\cite{MOADK} rates are used to determine the ionization rates and ionization times. Time-dependent Schr\"odinger equation (TDSE) is then solved for $H_2^+$ in chirped laser fields. For a preliminary understanding of the chirp effect on CEP-controlled dissociation, we first investigated the two-level model for the dissociation of $H_2$ with chirped laser pulses,  
\begin{gather}
\begin{align}
    &\begin{bmatrix}
        i\dot{\psi_g}(R,t) \\
        i\dot{\psi_u}(R,t)
    \end{bmatrix} \nonumber\\
    &= \begin{bmatrix}
        -\nabla^2/2\mu_p+V_g(R) &  -\mu(R)E(t) \\
        -\mu(R)E(t) & -\nabla^2/2\mu_p+V_u(R)  
    \end{bmatrix}
    \begin{bmatrix}
        \psi_g(R,t) \\
        \psi_u(R,t)
    \end{bmatrix},
\end{align}
\end{gather}in which we consider a perturbative chirping. Here, $\mu_p$ is the reduced mass of the nuclei. For a clear understanding, we take the initial wave-packet to be at the electronically excited $ungerade$ state, since, otherwise a large residue of the initial state would remain in the ground state~\cite{Vrakking2011,chu2004beyond}. To determine the interstate transitions and dynamical origin of these transitions (i.e., adiabatic, diabatic, or mixed dynamics), we calculate the avoided-crossing passage probability by the Landau-Zener formula as $P_{diabatic}(R,t)=\exp\{-\pi(V_u(R)-V_g(R))^2/(4\omega_0\mu(R)E(t))\}$ where $R$, $V_u$, $V_g$, $\omega_0$, $\mu(R)$ and $E(t)$ are intermolecular distance, ungerade and gerade states potentials, angular frequency of pulse, $R$-dependent transition dipole moment and time-dependent laser field, $  E(t)=E_0\exp(-4ln(2)\frac{t^2}{\tau^2})\cos(\omega_0 t + \phi_0 + \frac{4ln(2)\eta}{\tau^2}t^2),   \label{eqn:laser_field}$
respectively. $\phi_0$ is the CEP of the TL pulse and $\tau$ is the duration (at FWHM) of the pulse.
Here we define an asymmetry parameter, $A=(P_R-P_L)/(P_R+P_L)$, to quantify the directional electron localization, where $P_{R/L}$ corresponds to localization at the right/left nucleus. 
Fig.~\ref{fig:two_level} shows that, the adiabatic part finishes at $\sim2.5$ fs and diabatic part starts at $\sim6.2$ fs for the TL pulse. For positive(negative) chirp values, these happen earlier(later). For positive chirp, higher frequencies appear later in the falling edge, resulting in a more rapid change per cycle (see Fig.~\ref{fig:two_level}(a)).
Electronic transitions happen earlier, maximum transition probability per-cycle is reduced and localization is established in the left nucleus ($A<0$), after when the electron population freezes out (see Fig.~\ref{fig:two_level}(d)). For negative chirp, falling edge is at lower frequencies; the electronic transitions happen later and maximum transitions per-cycle are increased, resulting in an increased asymmetry in the mixed(adiabatic+diabatic) dynamics part ($i.e.$, within $~2-6$ fs). The electron localization is therefore established in the right nucleus ($A>0$). Regardless of the sign of chirp, absolute asymmetry is enhanced relative to that of the TL pulse and a linear variation of asymmetry with chirp in the form of $A\propto-\eta$ is found (see Fig.~\ref{fig:two_level}(d)).

\begin{figure*}[t!]
  \includegraphics[width=0.9\textwidth]{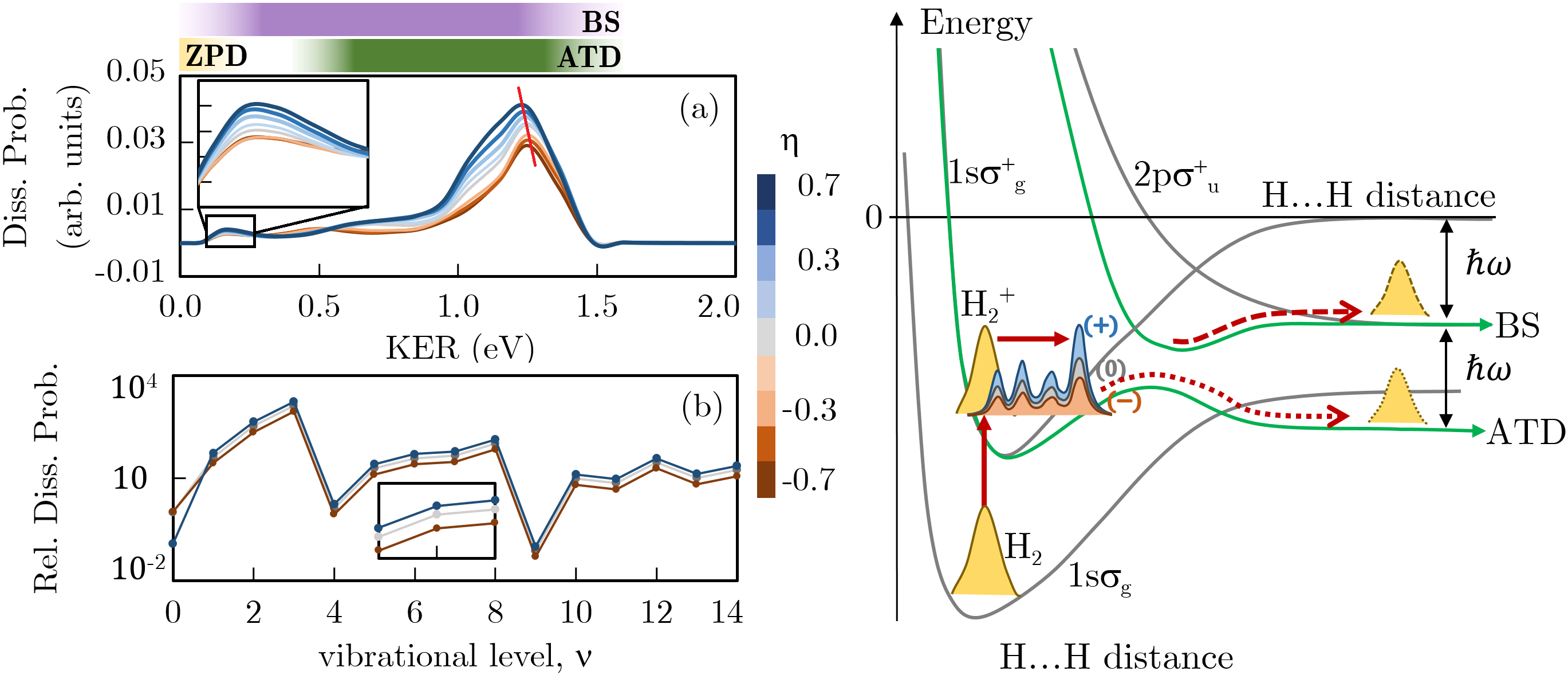}
  \caption{(color online) (a) variation of CEP-integrated KER spectra of dissociation by chirp. Inset: Zoom in to the 0.1-0.3 eV interval. (b) Contributions to the dissociation from different vibrational levels. Inset: Zoom in to the 5-7th vibrational states. (c) Energy curves of diabatic (grey) gerade and ungerade states (green) and possible BS and ATD dissociation channels. Diagram also illustrates how positive (+), transform-limited (0) and negative (-) chirp influence vibrational population transfer to the same vibrational state, based on (b). An 800nm Gaussian pulse with two-cycles at FWHM and I=150 TW/cm$^2$ is used.\\}
  \label{fig:amp_diss}
  \vspace{-8.0mm}
\end{figure*}

For a complete theoretical picture on the chirp effect in ultrashort dissociation of $H_2^+$, we solved Time Dependent Schr\"odinger Equation (TDSE) and obtain the Kinetic Energy Release (KER) of dissociation by chirped pulses to calculate CEP-dependence of dissociation and asymmetry. We use the TDSE defined by:
\begin{equation} \label{eq1}
\begin{split}
 i\frac{\partial}{\partial t} & \Psi (r,R,t)   =  [ T_e + T_n \\
& +  V_e(r,R) + V_n(R) + zE(t)  ] \Psi (r,R,t) . 
\end{split}
\end{equation} 
where $V_e(r,R)$ and $V_n(R)$ are the electron-nuclei, nucleus-nucleus, respectively, and last term is the linearly-polarized molecule-field interaction potential. We choose a two-cycles field at FWHM with wavelength $800$ nm and with an intensity of $150$ TW/cm$^2$. We use the same form of the chirped laser field, mentioned previously. We assume that $H_2^+$ is produced by on-the-fly ionization of the neutral hydrogen molecule, determined by MO-ADK, with Franck-Condon distributed vibrational levels on $1s\sigma_g$ potential curve and we examined different chirp values between $\eta=-0.7$ and $\eta=0.7$. In order to calculate the kinetic energy release (KER) spectra of the field-induced dissociation, i.e.~H$_2^+ + \gamma \rightarrow \mathrm{p}^++\mathrm{H}$, we employ the so-called ``virtual-detector" method  ~\cite{feuerstein2003}. In this approach, we first define $k(t)=m j(t)/\rho(t)$ as the time-dependent wave-packet momentum, calculated at point $R=R_d$, where we position the virtual-detectors at $R_d=10$ a.u.. $j(t)$ is the flux-density operator, $j(t)=\frac{1}{m} \mathrm{Re}[\langle\Psi(t)| \widehat{p}|\Psi(t)\rangle]|_{R=R_d}$, and $\rho(t)$ is the probability of finding the particle at $R=R_d$. $m$ is the mass of the particle. A binning/histogramming procedure is then necessary in order to derive the KER distribution $P(E)$ of H$_2^{+}$. The total KER of dissociation is determined from
\begin{equation}
    P_{tot} (E,\phi)\propto\int w_{ion}(t_i;\phi)P(t_i;E,\phi)dt_i,
\end{equation}where $w_{ion}(t_i;\phi)$ is the ionization rate of $H_2$ at instant $t_i$ and $P(t_i;E,\phi)$ is the dissociation probability of $H_2^+$ born at $t_i$. The integration is approximated in the form $\int dt \simeq \Sigma_n \delta(t-nT/2)$, since ionization is effectively high at the maxima of the pulse. 

There is a linear increase in dissociation probability by chirp which can be seen in Fig.~\ref{fig:amp_diss}(a). The increase is even more dramatic within the KER range of $1.0$-$1.3$ eV. The maximum of the distribution appears to be left shifted, consistent with the recent experiments for longer pulses~\cite{prabhudesai2010,natan2012}. To understand such variations, we have calculated the level-resolved contributions to dissociation and employed a Floquet-type analysis and used $E_{KER}=n \omega - |E_\nu|$, where $E_\nu$ is the vibrational energy and $n$ is the number of photons. Here, to eliminate the time-dependence of the laser frequency due to chirp, we considered a time-averaged frequency as, $\langle\omega\rangle = \omega_0 + \eta/\tau$ for the analysis of the results. Different dissociation channels (i.e., ZPD, BS, ATD, etc. with net $\nu = 0, 1, 2, ...$ absorption, respectively) contribute to dissociation. First, $\langle E_{KER} \rangle=\nu [\omega_0 + \eta/\tau] - |E_\nu|$, in fact, enounce that the kinetic-energies of the dissociating wave-packet is independent of chirp, since higher(lower) vibrational levels would $in$ $turn$ contribute to KER if positive(negative) chirp is used. Therefore, the variation of KER spectra with chirp is caused by the variations in vibrational excitations. To support this, we calculated the chirp-dependence of the dissociating vibrational population rates based on the projections of the dissociating wave-packet to each vibrational state, $\nu$, for positive, TL and negative chirp values. As shown in Fig. 2(b), and with the help of the diagram in Fig. 2(c), we find that the vibrational levels are coherently de-populated with positive(negative) chirp relative to the TL pulse. Its effect is more pronounced within the KER range $1.0$-$1.3$ eV, since both $\nu > 9$ rendering the BS channel and first few $\nu$ rendering the ATD channel (of which are at high populations) collectively contribute this KER range (see Fig. 2(a)-(c)). To conclude, the variation of KER in the dissociation probability is directly related to the variations in vibrational population transfer by chirp and to the dissociation channel.

\begin{figure}[t!]
  \includegraphics[width=\linewidth]{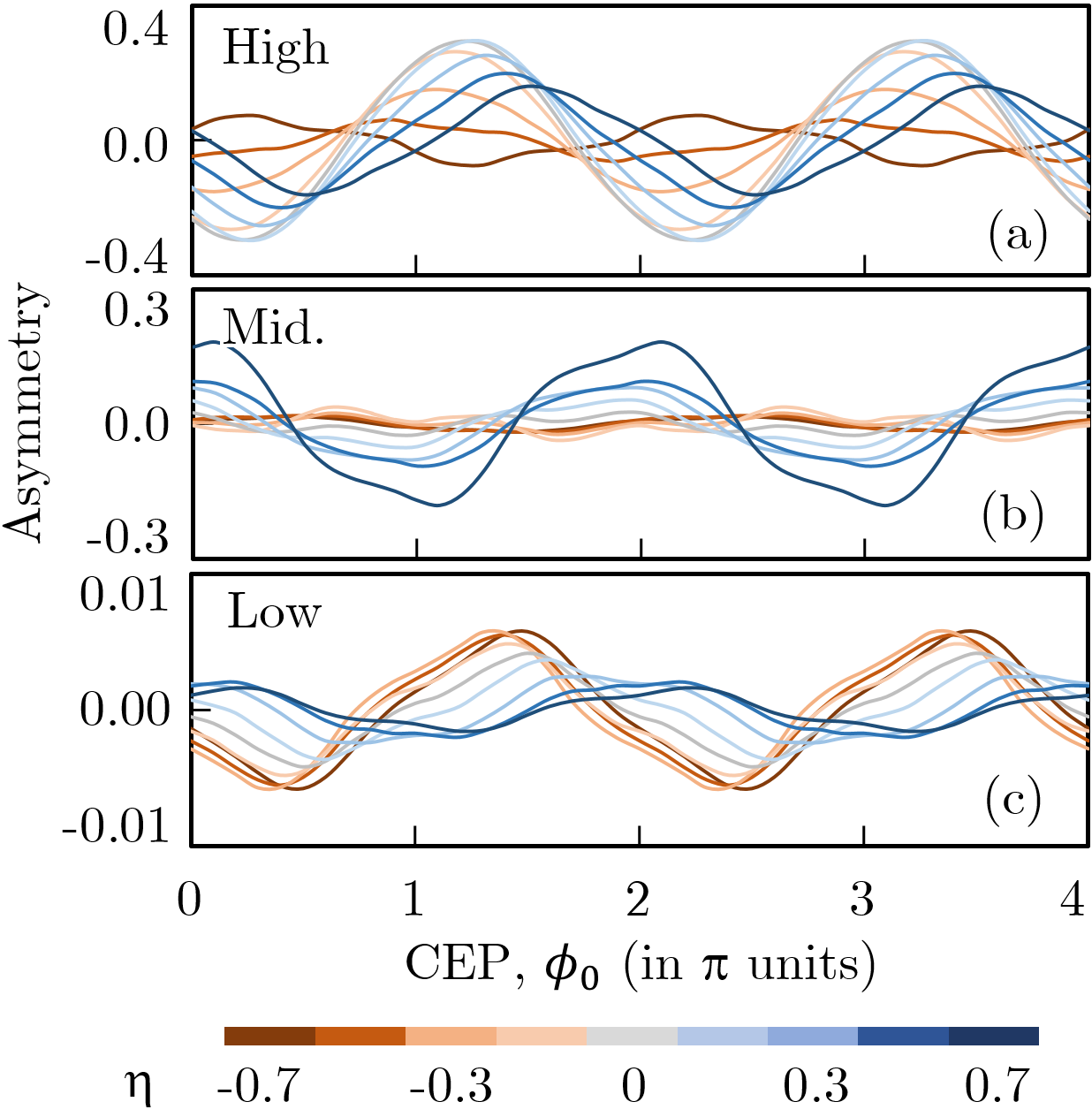}
  \caption{(color online) KER-integrated asymmetry for (a) high ($1.3-1.5$ eV) (b) middle ($0.8-1.0$ eV) and (c) low ($0.1-0.3$ eV) energy regions. CEP variation shifts to the left or right depending on the energy region linearly with chirp, consistent with the first order approximation of the impact of chirped CEP on asymmetry. Laser parameters are the same as in Fig.~\ref{fig:amp_diss}.\\}
  \label{fig:avg_asyms}
  \vspace{-8.0mm}
\end{figure}

\begin{figure*}[t!]
  \includegraphics[width=0.9\textwidth]{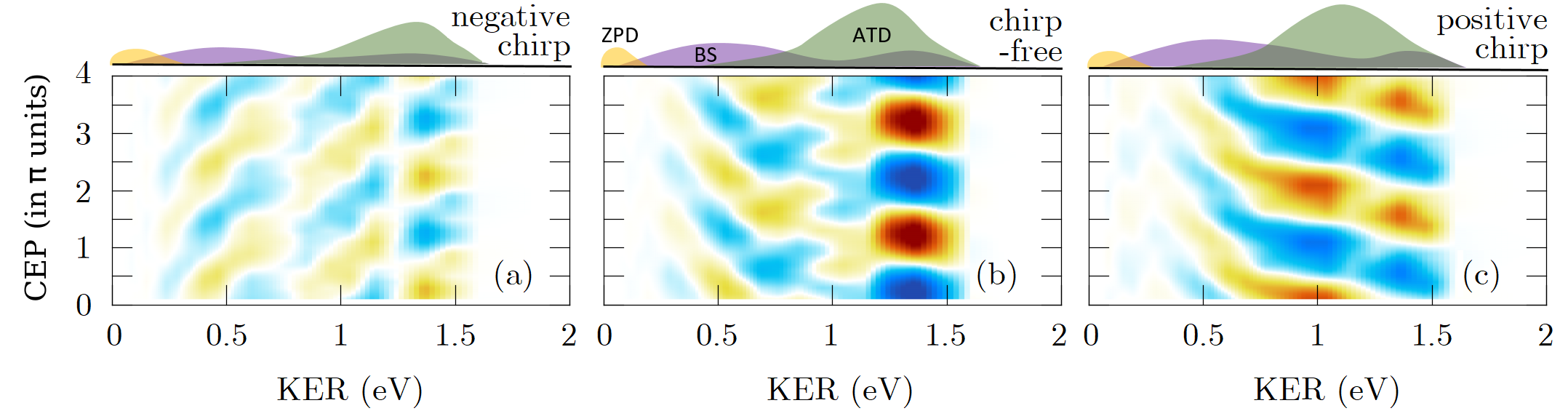}
  \caption{(color online) Asymmetry maps for (a) negative chirp ($\eta=-0.7$) (b) transform limited case ($\eta=0$) and (c) positive chirp ($\eta=0.7$) cases. Top panel: Contributions of different dissociation channels to KER determined from Fig.(\ref{fig:amp_diss}).\\}
  \label{fig:asyms}
  \vspace{-8.0mm}
\end{figure*}

We next calculate the CEP-dependence of dissociation probability and electron-localization asymmetry in ultrashort chirped laser pulses. The energy-resolved electron-localization asymmetry $A(E,\phi)$ is computed from $A=(P_R-P_L)/(P_R+P_L)$ with the help of Eq. 3, where $P_{R/L}(E,\phi)$ correspond to dissociation through the right/left and where $P_{tot}=P_R+P_L$. Chirped CEP is indeed time-dependent having the analytical form $\phi(t,\eta)=\phi_0+\eta t^2/\tau^2$, where $\eta t^2/\tau^2$ can be considered as the instantaneous phase variation in CEP, $i.e.$ ``phase-of-phase" \cite{bauerPP2015}. To comprehend how chirp affects the CEP dependency of dissociation probability and asymmetry, we time-averaged the chirped CEP as a first-order approximation and found that $\langle\phi(t,\eta)\rangle = \phi_0 + 4\eta/3$. This indicates that the phase-of-phase typically varies linearly by $4\eta/3$ with chirp.  We calculate the KER-CEP maps of the dissociation and asymmetry for various chirp values and present CEP variation of asymmetry results with KER integration over different energy ranges (see Fig. 3). One can see an apparent chirp modulation in phase in the CEP-dependent asymmetry. For high-energy region the amplitude of asymmetry of dissociation linearly decreases and CEP-dependence right-shifts for both positive and negative chirps. For the mid-energy region the amplitude linearly enhances and the CEP-dependence right-shifts from negative upto positive chirp. Finally, in contrast to the mid-energy case, for low-energy region the amplitude of asymmetry reduces and right-shifts. Specifically, a linear-fitting of the $\eta$ variation in the CEP positions of the maxima for the $1.3$-$1.5$ eV region, $0.8$-$1.0$ eV region and $0.1$-$0.3$ eV region (see Fig~\ref{fig:avg_asyms}) gave a slope of $1.21\pm0.09$, $1.40\pm0.12$ and $1.92\pm0.45$, respectively, which are consistent with 4/3 ($\simeq 1.33$) in the first-order approximation for CEP, $i.e.$ $\langle\phi(t,\eta)\rangle = \phi_0 + 4\eta/3$. Based on these considerations, one can, in general, approximate the CEP variation of asymmetry for a certain chirp in the form of $A(\phi,\eta) \simeq A_0(\eta)cos(\phi)$, where $\phi=\phi(\eta)$ (see Fig \ref{fig:avg_asyms}). 

Even though there is a linear relationship between the chirp and the CEP-dependence of asymmetry, it is clear that the mechanism behind the right-left shifts and the variation in the amplitudes for different chirp values and for different KER regions is much more complicated. But, apparently, from our analysis of the dissociation probability and its relation to the dissociation pathways show that they are playing the central role here. 
Besides, in general, interferences between states with different parity, e.g. between gerade and ungerade states in case of $H_2^+$, are needed to achieve asymmetry in electron localization \cite{esry}. Such superpositions can be reached by interferences of different paths such as zero-photon dissociation (ZPD) and bond softening (BS) or BS and above threshold dissociation (ATD). However, as in our case, this asymmetry can also be established in a single dissociation path with varying photon-transition energies in chirped pulses.
The ZPD-BS interference and BS-ATD interference produce asymmetry in the low and the remaining (mid and high) regions, respectively. 
For the low KER ($0.1$-$0.3$ eV), $\nu \simeq 5$-$7$ levels contribute to BS and very high $\nu$ values contribute to ZPD in ultrashort pulses, but their probabilities are low (see Fig. 2(b)). The decreasing amplitude in the asymmetry from negative to positive $\eta$ observed for the low KER in Fig. 3(c) is due to the systematic increase in the dissociation from BS relative ZPD in this KER region. In other words, increasing dissociation from BS relative to ZPD results in an overall decrease in the overlap of ZPD-BS KER regions. This is also evident from the systematic increase in the dissociation in the low KER region shown by the inset of Fig. 2(a). For the high KER range ($1.3$-$1.5$ eV) the asymmetry of the negative chirp decreases relative to TL pulse, which can be attributed to the decreasing dissociation rate for both BS and ATD in this region. The decrease in the asymmetry for the positive chirp $-$ even though dissociation rate increases $-$ is due to the shift of ATD to lower KER region, resulting in reducing BS-ATD overlap. These are also in compliance with increasing asymmetry around mid KER region $0.8$-$1.0$ eV for positive chirp as well as weak asymmetry for the negative chirp in the same region relative to the TL pulse. 

In conclusion, we have investigated the chirp effect on CEP-controlled dissociation and asymmetry and found that exploiting the chirp modulates electron localization during dissociation. Dissociation is enhanced by chirp and CEP dependence of both the dissociation and the asymmetry shift with the chirp parameter. We find that ZPD, BS and ATD pathways are selectively induced depending on the sign and value of chirp and their interferences alternatively contribute to the KER spectra. We find a linear relationship between the chirp and dissociation probability. We anticipate that our results for the chirped pulses provide an alternative perspective in the sub-cycle control of electron dynamics in molecular reactions.   

S.A.K. and I.Y. are grateful for the support by the Research Fund of Marmara University, Project Number FEN-C-DRP-230119-0005. S.A.K. is partially supported by the Turkish Funding Agency, TUBITAK, by the 2211 Graduate Bursary Program. P.R. and M.F.K. are grateful for support by the German Research Foundation via KL1439/11-1 and the center of excellence "Munich Centre of Advanced Photonics". This work was supported by the project by the project Advanced research using high intensity laser produced
photons and particles (No. CZ.02.1.01/0.0/0.0/16\_019/0000789) from European Regional Development Fund (ADONIS). Computing resources for simulations used in this work were provided by the National Center for High Performance Computing of Turkey (UHeM) under grant number 5005902018.
Data analysis are performed at the Simulations and Modelling Research Lab (Simulab), Physics Department of MU.

\bibliographystyle{apsrev4-1}
\bibliography{main}

\end{document}